\documentclass[aps,pre,superscriptaddress,reprint]{revtex4-2}
\usepackage{graphicx}
\usepackage{amsmath}
\usepackage{amsfonts}
\usepackage[utf8]{inputenc}
\usepackage{bm}
\usepackage{hyperref}
\hypersetup{colorlinks}
\hypersetup{citecolor=blue}

\DeclareMathOperator{\Tr}{Tr}

\begin{document}
\title{Stokes drag on a sphere in a three-dimensional anisotropic porous medium}
\author{Andrej Vilfan}
\email{andrej.vilfan@ijs.si}
\affiliation{Jožef Stefan Institute, 1000 Ljubljana, Slovenia}
\author{Bogdan Cichocki}
        \email{bogdan.cichocki@fuw.edu.pl}
    \affiliation{Institute of Theoretical Physics, Faculty of Physics, University of Warsaw, Pasteura 5, 02-093 Warsaw, Poland}
\author{Jeffrey C. Everts}
        \email{jeffrey.everts@fuw.edu.pl}
    \affiliation{Institute of Theoretical Physics, Faculty of Physics, University of Warsaw, Pasteura 5, 02-093 Warsaw, Poland}
        \affiliation{Institute of Physical Chemistry, Polish Academy of Sciences, 01-224 Warsaw, Poland}    
\date{\today}

\begin{abstract}
We study the hydrodynamic drag force exerted on a sphere in a static anisotropic porous medium. This problem is analysed using the Brinkman-Debye-Bueche equations with an axisymmetric shielding (or permeability) tensor. Using the exact Green's functions for this model fluid within a single-layer boundary element formulation, we numerically compute the friction tensor for a translating sphere subjected to stick boundary conditions. Furthermore, we derive approximate analytical expressions for small anisotropy using the Lorentz reciprocal theorem. By benchmarking this result against the numerical solutions, we find that a linear approximation is valid in a broad parameter regime. Our results are important for studying self-diffusion in general anisotropic porous media, but can also be applied to small tracers in nematic fluids composed of disk- or rod-like crowders.
\end{abstract}
\maketitle

\section{Introduction}

In soft matter and biology, liquids permeating porous media are often found due to the occurrence of (polymeric) fibres, membranes, and colloidal particles \cite{Kuhn:2011, Boon:2012, Charras:2013, Sasaki:2017, Lauga:2025}. It is a key challenge to describe the flow through such systems -- typically at a low Reynolds number--while also accounting for the precise porous microstructure of the medium \cite{Raccis:2011, Blunt:2013, Arezoo:2018}. Therefore, understanding the macroscale flow properties often requires a coarse-grained approach. In the Brinkman-Debye-Bueche (BDB) method \cite{Brinkman:1949, Debye:1948}, the porous features of the system are represented by an effective mesoscopic medium that exerts an additional force density on the permeating fluid that is linear in the fluid velocity. Although this method has been widely studied \cite{Brady:1987} and applied \cite{Feng:1998, Uematsu:2015, Jezewska:2024}, mostly isotropic effective media have been considered. Despite the possible profound effects of anisotropy \cite{Yang:1999, Dettmer:2014}, there are only a few results available for BDB models where the pores or building blocks are, on average, aligned in a specific direction. Here, we want to address this issue in the context of self-diffusion of a spherical tracer.

The main motivation for studying this problem is a series of experiments by Kang \emph{et al.} \cite{Kang:2005, Kang:2006, Kang:2007}, where the long-time self-diffusion coefficient of a spherical tracer particle is determined in a network of nematically ordered rod-like particles. These experiments have been interpreted using effective medium theories based on cluster expansions incorporating rod-sphere hydrodynamic interactions \cite{Kang:2005, Kang:2006, Kang:2007, Guzowski:2008, Cichocki:2009}, which are only valid for low volume fraction of rods on the level of the friction tensor. Furthermore, only isotropic hydrodynamic screening was included in the model, even in cases where the network of rods is nematic.
In these works, the spherical tracers are much smaller than the rod-like crowders and are thus not expected to distort the nematic host. Although confinement and crowding do not necessarily influence diffusion in the same manner \cite{Conrad:2016}, we can approximate the crowded environment of rods as an anisotropic porous medium for small enough tracer sizes.  Therefore, an anisotropic BDB approach gives a different perspective on this problem and could correct the theories presented in Refs. \cite{Kang:2005, Kang:2006, Kang:2007, Guzowski:2008, Cichocki:2009}. 

A direct way to incorporate anisotropy in the BDB approach is to replace the shielding length (which is equivalent to the permeability) with a shielding tensor that reflects the spatial symmetry of the underlying porous mass \cite{bear}. For this case, the Green's functions have been analytically computed in two spatial dimensions \cite{Kohr:2007} and later generalised to three spatial dimensions for axisymmetric systems up to one integral \cite{Cichocki:2010}. Such analytical expressions are important because they can be directly used in the Boundary Element Method (BEM) \cite{Pozrikidis:2002} to compute, for example, hydrodynamic drag forces. 

This work focuses on the numerical and analytical computation of the Stokes drag force on a translating sphere suspended in an unbounded axisymmetric BDB fluid. 
The paper is organised as follows. In Sec.\ \ref{sec:problem}, we describe the equations governing the flow in an anisotropic BDB fluid and present the corresponding Green's functions. Furthermore, we formulate the friction problem for a sphere in anisotropic BDB fluid, which is relevant for self-diffusion and hydrodynamic drag studies. In Sec.\ \ref{sec:ana}, we analyze the friction problem for small anisotropy with the help of the Lorentz reciprocal theorem. In Sec.\ \ref{sec:num}, we compute the friction tensor for arbitrary anisotropy using the BEM and compare the results with the analytical approximation of Sec.\ \ref{sec:ana}. We present our conclusions and an outlook in Sec.\ \ref{sec:con}.

\section{Description of the model fluid}
\label{sec:problem}
We consider an incompressible, anisotropic BDB fluid, described by the (average) fluid velocity ${\bf v}({\bf r})$ and pressure $p({\bf r})$. Conservation of linear momentum and the incompressibility condition, respectively, give
\begin{gather}
\eta_\mathrm{eff}\nabla^2{\bf v(r)}-\eta\boldsymbol{\kappa}^2\cdot{\bf v(r)}-\nabla p({\bf r})=-{\bf f(r)}, \label{eq:anisodbb1}\\ \nabla\cdot{\bf v(r)}=0\label{eq:anisodbb2}.
\end{gather}
Here, $\eta$ is the shear viscosity of the fluid in the porous mass, $\eta_\mathrm{eff}$ is the effective viscosity of the medium composed of the porous mass and the fluid, ${\bf f(r)}$ is a force density acting on the fluid, and $\boldsymbol{\kappa}$ is a spatially constant shielding tensor. Furthermore, we consider cylindrical (or ``nematic'') symmetry of $\boldsymbol{\kappa}$,
\begin{equation}
\boldsymbol{\kappa}=\kappa_\parallel\hat{\bf n}\hat{\bf n}+\kappa_\perp(\bm{\mathsf{I}}-{\hat{\bf n}\hat{\bf n}}), \label{eq:shielding}
\end{equation}
with the axis of cylindrical symmetry $\hat{\bf n}$. Here, $\kappa_\parallel$ and $\kappa_\perp$ are inverse ``screening'' lengths parallel to $\hat{\bf n}$ and in the plane perpendicular to $\hat{\bf n}$, respectively. Eqs.~\eqref{eq:anisodbb1} and \eqref{eq:anisodbb2} are valid for a stationary porous mass.

The second term in Eq.~\eqref{eq:anisodbb1} can be interpreted as an external force density exerted by the porous mass on the fluid. The precise form of this (Brinkman) damping force is chosen so that for small velocity gradients (or $\eta_\mathrm{eff}=0$) and $\bf f=0$, the Darcy law is retrieved \cite{Darcy, Whitaker:1986}. Consequently, the inverse of $\boldsymbol{\kappa}^2$ can be interpreted as the permeability tensor of the porous mass. In this work, we take $\eta_\mathrm{eff}=\eta$. Cases where $\eta_\mathrm{eff}\neq\eta$ and non-zero can be taken into account by a suitable rescaling of $\boldsymbol{\kappa}$ depending on the problem of interest.

\begin{figure*}
\centering
\includegraphics[width=\textwidth]{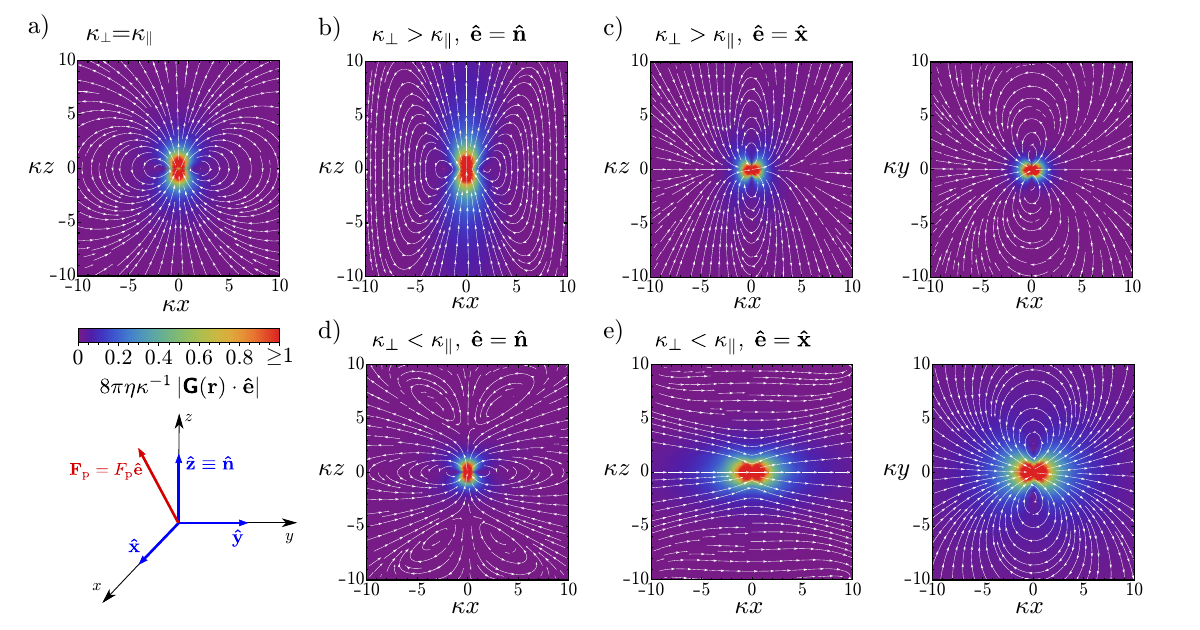}
\caption{Streamlines and dimensionless magnitude of the fluid velocity field ${\bf v(r)}=\bm{\mathsf G}({\bf r})\cdot{\bf F}_\mathrm{p}$ in response to a point force density with strength ${\bf F}_\mathrm{p}=F_\mathrm{p}\hat{\bf e}$. The coordinate system and corresponding unit vectors are defined in the lower left panel. (a)~Velocity in an isotropic medium. (b,c) Velocity in a medium with stronger transverse screening, $\kappa_\perp=5 \kappa_\parallel$. (d,e) Velocity in a medium with stronger longitudinal screening $\kappa_\perp=(1/5)\kappa_\parallel$ (d,e). Panels (a), (b) and (d) show the response to a parallel force, $\mathbf{\hat e}={\bf{\hat n}}$ and panels (c) and (e) show the response to a perpendicular force $\mathbf{\hat e}= {\bf\hat x}$ in two different planes. Secondary eddies can appear in the case $\kappa_\parallel> \kappa_\perp$ and longitudinal force (d).}
\label{fig:green}
\end{figure*}
\subsection{Point force response}
The fundamental solution can be obtained by setting ${\bf f(r)}={\bf F}_\mathrm{p}\delta({\bf r})$. By linearity of Eqs.~\eqref{eq:anisodbb1} and \eqref{eq:anisodbb2}, the solution can be written as ${\bf v(r)}=\bm{\mathsf{G}}({\bf r})\cdot{\bf F}_\mathrm{p}$ and $p({\bf r})={\bf Q(r)}\cdot{\bf F}_\mathrm{p}$, which defines the Green tensor $\bm{\mathsf{G}}({\bf r})$ and pressure vector ${\bf Q(r)}$. Their expressions are known up to one integral to be numerically evaluated, see Ref.~\cite{Cichocki:2010}. In a cylindrical coordinate system $(\rho,\phi,z)$ where $\hat{\bf n}$ is taken to be the $z$ axis, we parametrize the fundamental solution as
\begin{gather}
4\pi\eta\bm{\mathsf{G}}({\bf r})=A(\rho,z)\hat{\boldsymbol{\rho}}\hat{\boldsymbol{\rho}}+B(\rho,z)(\hat{\boldsymbol{\rho}}\hat{\bf n}+\hat{\bf n}\hat{\boldsymbol{\rho}})+C(\rho,z)\hat{\bf n}\hat{\bf n} \nonumber \\
+D(\rho,z)\hat{\boldsymbol{\phi}}\hat{\boldsymbol{\phi}} \label{eq:funG}
\end{gather}
and
\begin{gather}
4\pi{\bf Q}({\bf r})=R(\rho,z)\hat{\boldsymbol{\rho}}+Z(\rho,z)\hat{\bf n} \label{eq:funQ}.
\end{gather}
The precise form of the scalar functions $A$, $B$, $C$, $D$, $R$, and $Z$ can be found in Ref.~\cite{Cichocki:2010}, however, for completeness, we also present these functions in Appendix \ref{sec:ApA}.

Despite the complicated form of these scalar functions, the short- and long-distance behaviour of the Green's functions have a simple form \cite{Cichocki:2010}. For short distances from the origin, viscous stresses are much larger than the Brinkman force density, and we find the Oseen (or Stokeslet) form
\begin{equation}
\bm{\mathsf{G}}({\bf r})\sim\frac{1}{8\pi\eta r}(\bm{\mathsf{I}}+\hat{\bf r}\hat{\bf r}), \quad \rho, z \ll \mathrm{min}(\kappa_\perp^{-1},\kappa_\parallel^{-1}). \label{eq:short}
\end{equation}
In contrast, for large distances the viscous stresses can be neglected. From the balance of the pressure gradient and the Brinkman damping force density, we find the expression of a stretched source dipole
\begin{equation}
\bm{\mathsf{G}}({\bf r})\sim\frac{\kappa_\parallel\kappa_\perp^2}{4\pi\eta} \left(\frac{3{\bf r}{\bf r}}{\bar{r}^5}-\frac{1}{\bar{r}^3}\boldsymbol{\kappa}^{-2}\right), \,\, \rho, z \gg \mathrm{max}(\kappa_\perp^{-1},\kappa_\parallel^{-1}), \label{eq:long}
\end{equation}
with scaled coordinate $\bar{r}=\sqrt{(\kappa_\perp\rho)^2+(\kappa_\parallel z)^2}$. Note that in contrast to the electrostatic case, the far-field behaviour is not screened, which is a direct result of the incompressibility condition. This is most transparent in the isotropic case, $\kappa_\parallel=\kappa_\perp$, for which closed-form expressions (without integrals) exist for the Green's functions. Here, the tensorial components of $\bm{\mathsf{G}}({\bf r})$ --defined as the functions $h_1$ and $h_2$ in Appendix \ref{sec:ApA}-- each contain one term that does not decay exponentially \cite{Kim} and have to be compensated by pressure.

Each velocity field due to a point source can be understood using Eqs.~\eqref{eq:short} and \eqref{eq:long} as limiting situations. Examples of the velocity fields ${\bf v}({\bf r})$ for different screening anisotropies and different directions of the force ${\bf F}_\mathrm{p}$ are shown in Fig.~\ref{fig:green}. Whereas the near-field solutions ($\kappa r \ll 1$) always correspond to a Stokeslet and the far field to a rescaled source dipole, the intermediate regime can take a complex form.  For a ratio $\kappa_\parallel/\kappa_\perp \gtrsim 3.324$, a pair of secondary vortex rings appears. They are reminiscent of the Moffatt eddies in a corner \cite{Moffatt1964} or in a cavity \cite{Higdon1985}, as well as toroidal eddies in a pipe \cite{blake79}. However, in all these examples flow recirculation is caused by confinement and the eddies in a pipe can even disappear in the presence of a BDB fluid with sufficiently strong screening \cite{DaddiMoussaIder.Vilfan2025}. The appearance of secondary eddies without confinement is therefore a unique feature of an anisotropic BDB fluid.

\subsection{Friction problem of a sphere suspended in an anisotropic porous medium}
\label{sec:anisofric}
We consider a sphere of radius $a$ moving with translational velocity ${\bf U}$ in a fluid-containing anisotropic porous mass. Since we are interested in the steady motion of the particle, it suffices to consider a sphere in the origin of the laboratory frame. For the corresponding friction problem, we need to solve the boundary value problem given by Eqs.~
\eqref{eq:anisodbb1} and \eqref{eq:anisodbb2} for $r>a$ and ${\bf f=0}$, supplemented by the boundary conditions ${\bf v}(a\hat{\bf r})={\bf U}$, and ${\bf v(r)}\rightarrow 0, \, p({\bf r})\rightarrow p_\infty$ (=constant) for $r\rightarrow\infty$. From ${\bf v(r)}$ and $p({\bf r})$ the corresponding hydrodynamic stress tensor can be computed: $\boldsymbol{\sigma}=-p({\bf r})\bm{\mathsf{I}}+\eta\{[\nabla{\bf v}({\bf r})]+[\nabla {\bf v}({\bf r})]^\mathrm{T}\}$, with T denoting the transpose. The force that the particle exerts on the fluid is given by
\begin{equation}
{\bf F}=\oint_{r=a}\boldsymbol{\sigma}({\bf r})\cdot d{\bf S}, \label{eq:force}
\end{equation}
with inward-pointing surface normal.
By linearity of Eqs.~\eqref{eq:anisodbb1} and \eqref{eq:anisodbb2} we write ${{\bf F}=\boldsymbol{\zeta}\cdot{\bf U}}$, where $\boldsymbol{\zeta}$ is the friction tensor. For our axisymmetric problem, it has the form
\begin{equation}
\boldsymbol{\zeta}=\zeta_\parallel\hat{\bf n}\hat{\bf n}+\zeta_\perp(\bm{\mathsf{I}}-{\hat{\bf n}\hat{\bf n}}), \label{eq:fric}
\end{equation}
where we defined the longitudinal (transverse) drag coefficient $\zeta_\parallel$ ($\zeta_\perp$) with respect to $\hat{\bf n}$.
Eq.~\eqref{eq:fric} is related to the diffusion tensor via the fluctuation-dissipation theorem, $\bm{\mathsf{D}}=k_\mathrm{B}T\boldsymbol{\zeta}^{-1}$, where $k_\mathrm{B}$ is the Boltzmann constant and $T$ temperature. Here, the components of $\bm{\mathsf{D}}$ describe diffusion along $\hat{\bf n}$ and in the plane perpendicular to $\hat{\bf n}$. The diffusion constant related to the mean-squared displacement is given by $\Tr(\bm{\mathsf{D}})/3$. In Secs.~\ref{sec:ana} and \ref{sec:num} we will determine $\boldsymbol{\zeta}$ analytically and numerically, respectively.

\section{Analytical results for small anisotropy}
\label{sec:ana}

In the case of small anisotropy, we will derive an approximate solution by means of the Lorentz reciprocal theorem \cite{Masoud.Stone2019}. The theorem provides an integral identity between the main problem, for which we choose an anisotropic BDB fluid, and the auxiliary problem, for which we take an isotropic BDB flow. In both problems, we treat the interaction of the fluid with the porous medium as an external force. In this case, the Lorentz reciprocal theorem for a sphere in an unbounded fluid reads \cite{Kim}
\begin{align}
&\oint_{r=a}{\bf v}_0({\bf r})\cdot[\boldsymbol{\sigma}({\bf r})\cdot d{\bf S}]-\int_{r>a} dV\, {\bf v}_0({\bf r})\cdot[\nabla\cdot\boldsymbol{\sigma}({\bf r})]= \nonumber\\
&\oint_{r=a}{\bf v}({\bf r})\cdot[\boldsymbol{\sigma}_0({\bf r})\cdot d{\bf S}]-\int_{r>a} dV\, {\bf v}({\bf r})\cdot[\nabla\cdot\boldsymbol{\sigma}_0({\bf r})], \label{eq:Lorentz}\end{align}
where ${\bf v(r)}$ satisfies Eqs.~\eqref{eq:anisodbb1}-\eqref{eq:shielding} with stress tensor $\boldsymbol{\sigma}({\bf r})$. The auxiliary problem has a velocity field ${\bf v}_0({\bf r})$ and stress tensor $\boldsymbol{\sigma}_0({\bf r})$ and is described by the same equations, but with isotropic shielding tensor $\boldsymbol{\kappa}_0=\kappa\bm{\mathsf{I}}$ (with $\kappa$ to be specified later). Furthermore, we impose the same boundary conditions (i.e.,  ${\bf v}(a\hat{\bf r})={\bf v}_0(a\hat{\bf r})={\bf U}$). 

The force that the sphere exerts on the fluid for the auxiliary system (given by ${\bf F}_0$) and the system of interest (given by ${\bf F}$) then satisfy the relation
\begin{equation}
{\bf U}\cdot({\bf F}-{\bf F}_0)=\eta \int_{r>a}dV\, {\bf v}_0({\bf r})\cdot(\boldsymbol{\kappa}^2-{\kappa}^2\bm{\mathsf{I}})\cdot{\bf v}({\bf r}) \label{eq:temp}
\end{equation}
where we used the boundary condition on the fluid velocity fields in Eq.~\eqref{eq:Lorentz}, that the divergence of the stress tensor gives the force density acting on the fluid, and that the considered shielding tensors are symmetric. In the isotropic auxiliary system, the friction tensor is analytically known, and we write ${\bf F}_0=\zeta_0({\kappa},a){\bf U}$, with $\zeta_0$ given by
\begin{equation}
\zeta_0({\kappa},a)=6\pi\eta a\left[1+{\kappa} a+\frac{({\kappa} a)^2}{9}\right]. \label{eq:isofric}
\end{equation}
Note the factor $1/9$ which is different from the result quoted by Brinkman \cite{Brinkman:1949}, which has a factor $1/3$ instead. This is a known property of the Brinkman equation: the drag of a moving sphere in a quiescent fluid is different from the drag of a stationary sphere in a fluid with constant ambient velocity \cite{Howells:1974, Anderson:1996}.
From Eq.~\eqref{eq:temp} we then find the exact relation
\begin{align}
&[\boldsymbol{\zeta}-\zeta_0({\kappa},a)\bm{\mathsf{I}}]:{\bf U}{\bf U}=\label{eq:Bogdan}\\
&\eta\Bigg[(\bar{\kappa}^2-\kappa^2)\bm{\mathsf{I}}+\frac{(\kappa_\perp^2-\kappa_\parallel^2)}{3}(\bm{\mathsf{I}}-3{\hat{\bf n}\hat{\bf n}})\Bigg]:\int_{r>a} dV\, {\bf v}_0({\bf r}){\bf v}({\bf r}),\nonumber
\end{align}
with $\bar{\kappa}^2=(2\kappa_\perp^2+\kappa_\parallel^2)/3$.
However, there is no analytical expression for ${\bf v}({\bf r})$. We obtain an approximate expression by setting $\kappa=(2\kappa_\perp+\kappa_\parallel)/3$ and choosing
\begin{equation}
\epsilon=\frac{\kappa_\perp-\kappa_\parallel}{\kappa}
\end{equation}
as a small parameter by using $\kappa_\perp={\kappa(1+\epsilon/3)}$ and $\kappa_\parallel=\kappa(1-2\epsilon/3)$. Our motivation for this specific choice of $\kappa$ is that then $\bar{\kappa}^2-\kappa^2={O}(\epsilon^2)$, and, therefore, we can neglect the first term between square brackets in the RHS of Eq.~\eqref{eq:Bogdan}. The second term after linearisation is of ${O}(\epsilon)$ and, consequently, for $\epsilon\rightarrow 0$ we can set ${\bf v}({\bf r})= {\bf v}_0({\bf r})$ in Eq.~\eqref{eq:Bogdan} for the remaining factor. We thus need to compute the dyadic integral
\begin{equation}
\int_{r>a}{d}V\, {\bf v}_0({\bf r}){\bf v}_0({\bf r})=c_1{\bf U}{\bf U}+c_2U^2\bm{\mathsf{I}}, \label{eq:dyad}
\end{equation}
where the structure of the integral follows from symmetry with $U=|{\bf U}|$. The coefficients $c_1$ and $c_2$ can be explicitly evaluated (see Appendix \ref{sec:ApB}). Note that $c_2$ does not contribute to the second term of the RHS in Eq.~\eqref{eq:Bogdan} due to the double contraction with a traceless tensor. We find to linear order in $\epsilon$,
\begin{equation}
\boldsymbol{\zeta}-\zeta_0(\kappa, a)\bm{\mathsf{I}}=6\pi\eta a\frac{63+2{\kappa}a}{270}(\kappa_\perp-\kappa_\parallel)a(\bm{\mathsf{I}}-3{\hat{\bf n}\hat{\bf n}})+o(\epsilon).
\label{eq:linearapprox}
\end{equation}
An interesting corollary of our result is that ${\Tr(\boldsymbol{\zeta}-\zeta_0\bm{\mathsf{I}})=o(\epsilon)}$, which suggests that there is a large parameter regime for which the diffusion constant is well approximated by the result from an isotropic analysis with $\kappa$ as inverse screening length. 

\begin{figure}
    \centering
    \includegraphics[width=0.8\columnwidth]{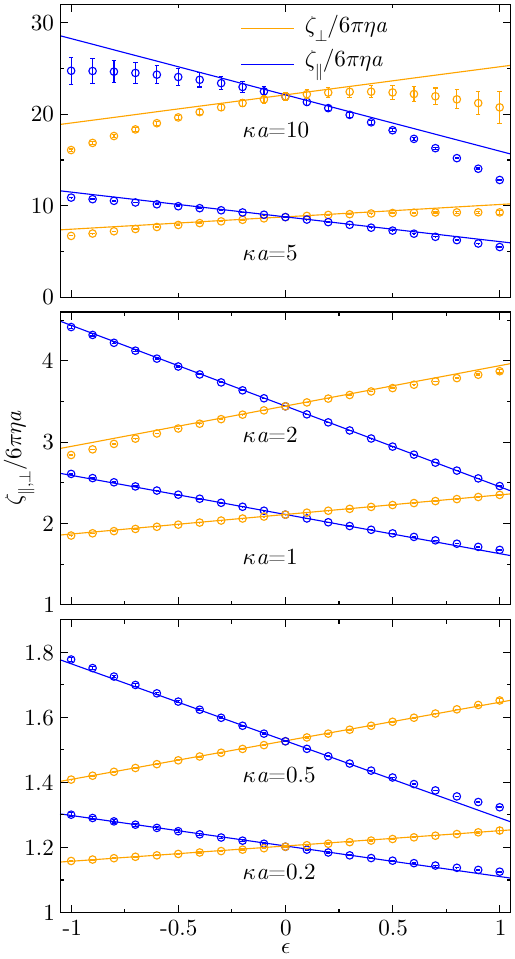}
    \caption{Transverse (orange) and longitudinal (blue) drag coefficients as a result of the linear approximation (Eq.~\eqref{eq:linearapprox}, solid lines) and numerical results using the boundary-element method (circles). Circles show the result with $N=2048$ boundary elements and the error bars the discrepancy to the result with $N=512$.}
    \label{fig:epsilonplot}
\end{figure}
\begin{figure}
    \centering
    \includegraphics[width=\columnwidth]{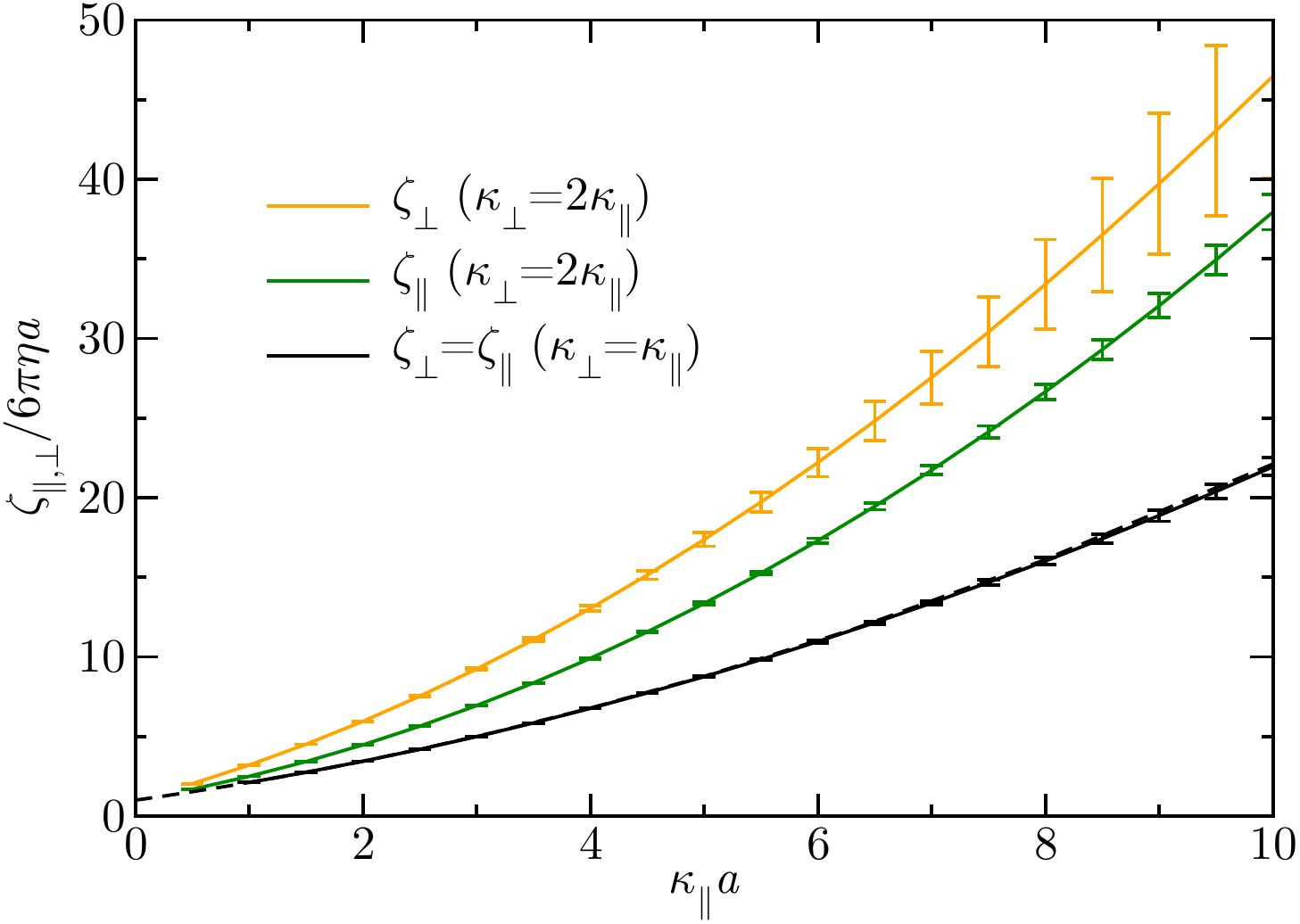}
    \caption{Transverse ($\zeta_\perp$) and longitudinal ($\zeta_\parallel$) drag coefficients, obtained using the boundary-element method ($N=2048$). The black line shows the result for an isotropic medium (with the dashed line showing the exact analytical result, Eq.~\eqref{eq:isofric}) and the orange/green lines for a medium with $\kappa_\perp=2\kappa_\parallel$. }
    \label{fig:zeta}
\end{figure}
\begin{figure*}
\centering
\includegraphics[width=0.95\textwidth]{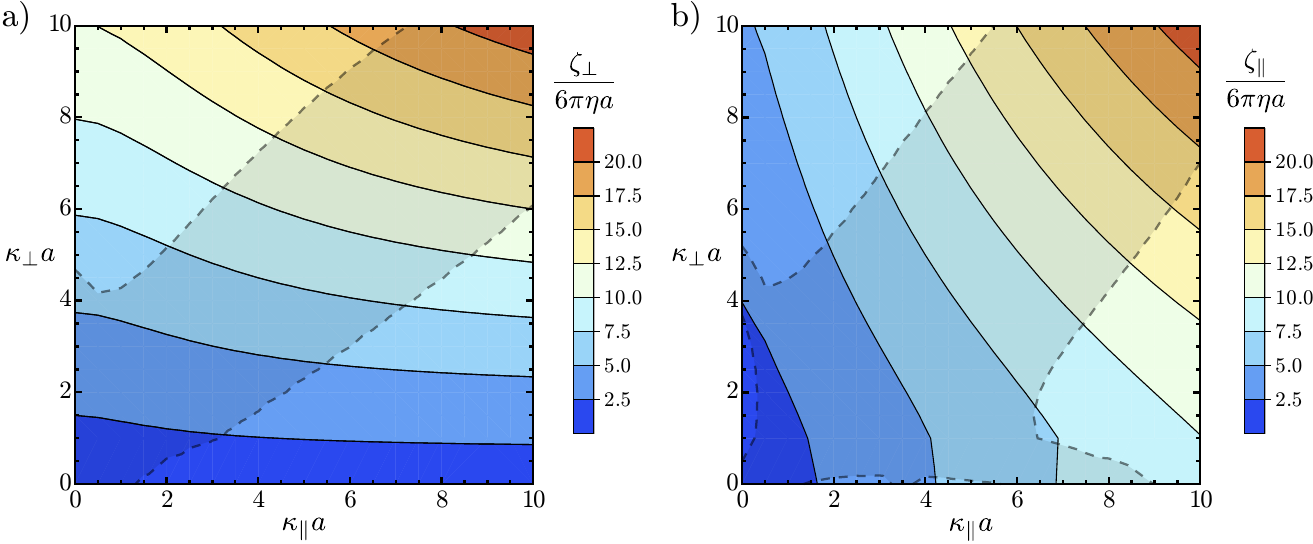}
\caption{Transverse (a) and longitudinal (b) drag coefficients, determined numerically using the BEM ($N=512$). The shaded region shows the parameter range for which the discrepancy between the linear approximation and the numerical result lies below $5\%$. }
\label{fig:contour}
\end{figure*}

A comparison between the linear approximation and full numerical results (see next section) is shown in Fig.~\ref{fig:epsilonplot}. Surprisingly, we see that our linear approximation Eq.~\eqref{eq:linearapprox} has a large range of applicability, even when $|\epsilon|$ is on the order of unity. Deviations with the numerical result only occur when $\kappa a$ is sufficiently large. We note that our asymptotic analysis in Eq. \eqref{eq:linearapprox} shows that the remainder term is of $o(\epsilon)$ for $\epsilon\rightarrow 0$. However, based on the numerical analysis, we expect that ${\bf v(r)}$ is an analytical function around $\epsilon=0$, and, therefore, we can safely hypothesize that the accuracy of Eq. \eqref{eq:linearapprox} is of $O(\epsilon^2)$. The precise regime where Eq.~\eqref{eq:linearapprox} is valid will be further analysed in Sec.~\ref{sec:num}.

\section{Numerical determination of the friction tensor}
\label{sec:num}

The numerical solution consists of numerically integrating the expressions for the Green's functions (Eqs.~\eqref{eq:a1}-\eqref{eq:a4} and \eqref{eq:a8}-\eqref{eq:a11}) and then solving the flow problem with the Boundary Element Method (BEM). We carried out the integration over $t$ or $s$ using the adaptive integration routine \texttt{QAGIU} from GNU Scientific Library (GSL). The code to compute the Green's functions is available in \cite{zenodo}. The obtained Green's functions were then used by a BEM solver using a single-layer formulation (adapted from the function \texttt{prtcl\_3d}, which is part of the BEMLIB package \cite{Pozrikidis:2002}). The surface of the spherical particle was discretised with 512 triangular elements. Finally, the obtained drag coefficients were reduced by the internal contribution, $(4/3)\pi \eta \kappa^2_{\parallel}a^3$ and $(4/3)\pi \eta \kappa^2_{\perp}a^3$ for the longitudinal and transverse drag, respectively, which is included in the result of a single-layer solution. Data in Figs.~\ref{fig:epsilonplot} and \ref{fig:zeta} were calculated with $4\times$ (i.e., 2048) the number of triangular elements and the difference between the two resolutions is shown as error bars.

The computed drag coefficients, $\zeta_\parallel$ and $\zeta_\perp$, are shown in Fig.~\ref{fig:zeta} for two values of the ratio $\kappa_\perp/\kappa_\parallel$. In the isotropic case ($\kappa_\perp=\kappa_\parallel$), the results show good agreement with the exact analytical expression (Eq.~\eqref{eq:isofric}). A minor deviation is visible at large values of $\kappa_\parallel a$, where the screening length becomes smaller than the size of a boundary element ($0.16a$ for 512 elements and $0.08a$ for 2048 elements). For the same reason, the accuracy deteriorates when $\kappa_\perp a > 10$. Our results show that in order to achieve an error of $\lesssim 1\%$, the mesh size needs to be chosen smaller than $1/\max(\kappa_\parallel,\kappa_\perp)$.

The whole range of numerical values of the drag coefficients $\zeta_\parallel$ and $\zeta_\perp$ is shown by the contour plots in Fig.~\ref{fig:contour}. Along with the linear plots in Fig.~\ref{fig:epsilonplot}, the results show that there is a wide parameter regime --indicated by the shaded region-- where the approximate result, given by Eq.~\eqref{eq:linearapprox}, agrees with the numerical value with a discrepancy smaller than $5\%$. 

\section{Concluding remarks}
\label{sec:con}
In this work, we have analysed the drag force on a translating spherical particle in a quiescent anisotropic BDB with axisymmetric shielding tensor. We numerically determined the components of the friction tensor by using the BEM, from which the diffusion tensor can be inferred. Furthermore, we have provided an approximate analytical formula that is valid in a wide range of the parameter space, which could be useful for experimentalists or as an input for theoretical models. The diffusion tensor considered in this work (which can be extracted from the friction tensor) is that of a sphere in a static anisotropic porous medium. Here, the porous mass itself does not exert a force on the tracer particle stemming from direct interactions, although there are indirect forces  mediated through the fluid. Furthermore, possible Brownian motion of the porous network is not included in our analysis. For the experiments listed in the introduction (Refs. \cite{Kang:2005, Kang:2006, Kang:2007}), our results should, therefore, be interpreted as a single-particle diffusion tensor in an effective medium. 

It would be interesting to expand the effective nematic theory analysed in Refs. \cite{Kang:2005, Kang:2006, Kang:2007, Guzowski:2008, Cichocki:2009} to include the effects of anisotropic hydrodynamic screening using the exact Greens function, and compare with our single-particle result.  We note that this approach was already proposed in Refs. \cite{Kang:2006, Kang:2007}, but at the time it was not possible due to the lack of solutions for anisotropic BDB fluids. Since the crowders are rod-shaped, the experiments correspond to the case where $\kappa_\perp>\kappa_\parallel$. The opposite case, $\kappa_\perp<\kappa_\parallel$, would correspond to a nematic system where the short axis of the particles are nematically ordered, which is the case, for example, for disc-like particles. However, certain experiments are essential for making a sensible comparison of theory with measurements. For example, we envisage that measurements of the diffusion coefficients for various tracer sizes are essential. Such studies are needed to assess whether the extracted screening lengths are independent of the probe size, as it is a property of the nematic medium. Such a consistency check is currently lacking for the existing experiments.

Beyond the scope of these experiments, our results have general merit for the study of anisotropic porous media within the BDB framework, generalizing the classical work by Brinkman \cite{Brinkman:1949} to the axisymmmetric case. Specifically, our results could be useful for problems where diffusion through porous media play an important role, like in the charging kinetics of supercapacitors \cite{Gupta:2024} or heterogeneous catalysis \cite{Gounder:2013}. Another possible direction is to investigate the interplay between anisotropic hydrodynamic screening with anisotropic electrostatic screening \cite{Everts:2021} for the diffusion or electrophoresis of charged particles, which is currently only touched upon for the isotropic case \cite{Kang:2007}.

\section*{Acknowledgements}
A.\ V. acknowledges funding from the Slovenian Research and Innovation Agency, grant no.\ P1-0099. 
J.~C.~E. acknowledges funding from the National Science Centre, Poland, within the SONATA BIS grant no.\ 2023/50/E/ST3/00452/R. 

\section*{Data availability}
The supporting data for this article are openly available from Zenodo \cite{zenodo}.
\newline
\appendix
\section{Scalar functions for parametrizing the fundamental solution}
\label{sec:ApA}
Here, we list the scalar functions defined in the parametrization of the fundamental solution, see Eqs.~\eqref{eq:funG} and \eqref{eq:funQ}. We need to distinguish the cases $\kappa_\perp>\kappa_\parallel$ and $\kappa_\perp<\kappa_\parallel$. Results are taken from Ref.~\cite{Cichocki:2010}, where only the cases for $z>0$ were presented. Here, we present expressions for general $z$. 

\begin{widetext}
\subsection*{\boldmath The $\kappa_\perp>\kappa_\parallel$ case}
The functions for $\kappa_\perp>\kappa_\parallel$ are given by
\begin{gather}
A(\rho,z)=\frac{1}{\kappa_\perp\rho^2}(e^{-\kappa_\perp |z|}-e^{-\kappa_\perp r})+\int_0^\infty dt\, \frac{t}{\Delta_\mathrm{b}(t)}\left[b_+(t)e^{-b_+(t)|z|}-b_-(t)e^{-b_-(t)|z|}\right]J_1'(t\rho), \label{eq:a1}\\
B(\rho,z)=-\mathrm{sgn}(z)\int_0^\infty dt\, \frac{t^2}{\Delta_\mathrm{b}(t)}\left[e^{-b_+(t)|z|}-e^{-b_-(t)|z|}\right]J_1(t\rho), \\
C(\rho,z)=-\int_0^\infty dt\, \frac{t^3}{\Delta_\mathrm{b}(t)}\left[\frac{e^{-b_+(t)|z|}}{b_+(t)}-\frac{e^{-b_-(t)|z|}}{b_-(t)}\right]J_0(t\rho), \\
D(\rho,z)=\frac{1}{r}e^{-\kappa_\perp r}-\frac{1}{\kappa_\perp\rho^2}(e^{-\kappa_\perp |z|}-e^{-\kappa_\perp r})+\int_0^\infty dt\, \frac{t}{\Delta_\mathrm{b}(t)}\left[b_+(t)e^{-b_+(t)|z|}-b_-e^{-b_-(t)|z|}\right]\frac{J_1(t\rho)}{t\rho}, \label{eq:a4}\\
R(\rho,z)=\frac{1}{2}\int_0^\infty dt\, \frac{t^2}{\Delta_\mathrm{b}(t)}\left\{[\Delta_\mathrm{b}(t)-2\kappa_\parallel^2+\kappa_\perp^2]\frac{e^{-b_+(t)|z|}}{b_+(t)}+[\Delta_\mathrm{b}(t)+2\kappa_\parallel^2-\kappa_\perp^2]\frac{e^{-b_-(t)|z|}}{b_-(t)}\right\}J_1(t\rho), \\
Z(\rho,z)=\frac{1}{2}\,\mathrm{sgn}(z)\int_0^\infty dt\, \frac{t}{\Delta_\mathrm{b}(t)}\left\{[\Delta_\mathrm{b}(t)-\kappa_\perp^2]e^{-b_+(t)|z|}+[\Delta_\mathrm{b}(t)+\kappa_\perp^2]e^{-b_-(t)|z|}\right\}J_0(t\rho),
\end{gather}
where a prime denotes differentiation with respect to the argument and $J_n$ is the $n$th-order Bessel function of the first kind. Furthermore, we have that
\begin{equation}
b_\pm(t)^2=t^2+\frac{1}{2}\left(\kappa_\perp^2\pm\sqrt{\kappa_\perp^4+4(\kappa_\perp^2-\kappa_\parallel^2)t^2}\right)\geq 0
\end{equation}
and $\Delta_\mathrm{b}(t)=b_+(t)^2-b_-(t)^2$.

\subsection*{\boldmath The $\kappa_\perp<\kappa_\parallel$ case}
The functions for $\kappa_\perp<\kappa_\parallel$ are given by
\allowdisplaybreaks
\begin{gather}
A(\rho,z)=-\frac{e^{-\kappa_\perp r}}{\kappa_\perp\rho^2}+\frac{2}{\pi}\int_0^\infty ds\, \frac{s^2\cos(sz)}{\Delta_\mathrm{c}(s)}\left[K_1'(c_+(s)\rho)-K_1'(c_-(s)\rho)\right], \label{eq:a8}\\
B(\rho,z)=-\frac{2}{\pi}\int_0^\infty ds\, \frac{s\sin(sz)}{\Delta_\mathrm{c}(s)}\left[c_+(s)K_1(c_+(s)\rho)-c_-(s)K_1(c_-(s)\rho)\right], \\
C(\rho,z)=\frac{2}{\pi}\int_0^\infty ds\, \frac{\cos(sz)}{\Delta_\mathrm{c}(s)}\left[c_+(s)^2K_0(c_+(s)\rho)-c_-(s)^2K_0(c_-(s)\rho)\right], \\
D(\rho,z)=\left(\frac{1}{r}+\frac{1}{\kappa_\perp\rho^2}\right)e^{-\kappa_\perp r}+\frac{2}{\pi}\int_0^\infty ds\, \frac{s^2\cos(sz)}{\Delta_\mathrm{c}(s)}\left[\frac{K_1(c_+(s)\rho)}{c_+(s)\rho}-\frac{K_1(c_-(s)\rho)}{c_-(s)\rho}\right], \label{eq:a11}\\
R(\rho,z)=\frac{1}{\pi}\int_0^\infty ds\, \frac{\cos(sz)}{\Delta_\mathrm{c}(s)}\left\{[\Delta_\mathrm{c}(s)-\kappa_\parallel^2]c_+(s)K_1(c_+(s)\rho)+[\Delta_c(s)+\kappa_\parallel^2]c_-(s)K_1(c_-(s)\rho)\right\}, \\
Z(\rho,z)=\frac{1}{\pi}\int_0^\infty ds\, \frac{s\sin(sz)}{\Delta_\mathrm{c}(s)}\left\{[\Delta_\mathrm{c}(s)+\kappa_\parallel^2-2\kappa_\perp^2]K_0(c_+(s)\rho)+[\Delta_\mathrm{c}(s)-\kappa_\parallel^2+2\kappa_\perp^2]K_0(c_-(s)\rho)\right\},
\end{gather}
where $K_n$ is the $n$th-order modified Bessel function of the second kind. Furthermore, we have defined the auxiliary functions
\begin{equation}
c_\pm(s)^2=s^2+\frac{1}{2}\left(\kappa_\parallel^2\pm\sqrt{\kappa_\parallel^4+4(\kappa_\parallel^2-\kappa_\perp^2)s^2}\right)\geq 0
\end{equation}
and $\Delta_\mathrm{c}(s)=c_+(s)^2-c_-(s)^2$.
\end{widetext}

\subsection*{\boldmath The $\kappa_\perp=\kappa_\parallel$ case}
The formulas in the case of anisotropic hydrodynamic screening can be explicitly evaluated for $\kappa_\perp=\kappa_\parallel$, for which the formulas for $\kappa_\perp<\kappa_\parallel$ and $\kappa_\perp>\kappa_\parallel$ give the same result. In this case, the fundamental solution takes the form
\begin{gather}
\bm{\mathsf{G}}({\bf r})=\frac{1}{4\pi\eta r}[h_1(\kappa r)\bm{\mathsf{I}}+h_2(\kappa r)\hat{\bf r}\hat{\bf r}] \label{eq:funGiso}
\end{gather}
and
\begin{gather}
{\bf Q}({\bf r})=\frac{1}{4\pi r^2}\hat{\bf r} \label{eq:funQiso}.
\end{gather}
Here, we defined
\begin{align}
h_1(x)&=-\frac{1}{x^2}+\left(1+\frac{1}{x}+\frac{1}{x^2}\right)e^{-x}, \\
h_2(x)&=\frac{3}{x^2}-\left(1+\frac{3}{x}+\frac{3}{x^2}\right)e^{-x}.
\end{align}
Note that all the expressions presented in this Appendix will give the dipolar form (Eq.~\eqref{eq:long}) in the far-field limit, as was shown in Ref.~\cite{Cichocki:2010}. However, there was a printing error in this manuscript. Eq.~65 in Ref.~\cite{Cichocki:2010} should read
\begin{equation}
C_{D}(\rho,z)=-\frac{\kappa_\perp^2}{\kappa_\parallel\bar{r}^3}+3\kappa_\perp^2\kappa_\parallel\frac{z^2}{\bar{r}^5}.
\end{equation}

\section{\boldmath Evaluation of coefficients $c_1$ and $c_2$}
\label{sec:ApB}

This Appendix provides the details for evaluating Eq.~\eqref{eq:dyad}. We parametrize the velocity field as
\begin{equation}
{\bf v}_0({\bf r})=[F(r)\bm{\mathsf{I}}+H(r)\hat{\bf r}\hat{\bf r}]\cdot{\bf U}, \label{eq:param}
\end{equation}
with 
\begin{gather}
F(r)=\frac{3}{2}\Bigg(\frac{\kappa ae^{\kappa a}}{(\kappa r)^3}\{[1+\kappa r+(\kappa r)^2]e^{-\kappa r}-1\}\nonumber \\
+\frac{a^3}{r^3}\frac{3e^{\kappa a}-3-3\kappa a-(\kappa a)^2}{3(\kappa a)^2}\Bigg)
\end{gather}
and
\begin{gather}
H(r)=\frac{3}{2}\Bigg(\frac{\kappa ae^{\kappa a}}{(\kappa r)^3}\{3-[3+3\kappa r+(\kappa r)^2]e^{-\kappa r}\}\nonumber \\
-\frac{a^3}{r^3}\frac{3e^{\kappa a}-3-3\kappa a-(\kappa a)^2}{(\kappa a)^2}\Bigg),
\end{gather}
as can be derived by directly evaluating the singularity form presented in Sec.\ 6.2.1 of Ref.~\cite{Kim}. Insertion of Eq.~\eqref{eq:param} into Eq.~\eqref{eq:dyad} gives $\int_{r>a} dV\, {\bf v}_0({\bf r}){\bf v}_0({\bf r})={{\bf U}\cdot\boldsymbol{\Lambda}\cdot{\bf U}}$, with rank 4 tensor
\begin{equation}
\boldsymbol{\Lambda}=\int_{r>a} dV\, [F(r)^2\bm{\mathsf{I}}\bm{\mathsf{I}}+F(r)H(r)(\bm{\mathsf{I}}\hat{\bf r}\hat{\bf r}+\hat{\bf r}\hat{\bf r}\bm{\mathsf{I}})+H(r)^2\hat{\bf r}\hat{\bf r}\hat{\bf r}\hat{\bf r}].
\end{equation}
Passing to spherical coordinates, the angular integrations can be performed with
\begin{gather}
\int_{\mathcal{S}^2}dS\, \xi_\alpha\xi_\beta=\frac{4\pi}{3}\delta_{\alpha\beta}, \\
\int_{\mathcal{S}^2}dS\, \xi_\alpha\xi_\beta\xi_\gamma\xi_\sigma=\frac{4\pi}{15}(\delta_{\alpha\beta}\delta_{\gamma\sigma}+\delta_{\alpha\gamma}\delta_{\beta\sigma}+\delta_{\alpha\sigma}\delta_{\beta\gamma}),
\end{gather}
with $[\hat{\bf r}]_\alpha=\xi_\alpha$ and $\mathcal{S}^2$ denoting the two-dimensional unit sphere. We can read off the coefficients by comparing them with Eq.~\eqref{eq:dyad}. We find
\begin{align}
c_1&=4\pi\int_a^\infty dr\, r^2\left[F(r)^2+\frac{2}{3}F(r)H(r)+\frac{2}{15}H(r)^2\right]\nonumber\\
&=\frac{a^2\pi(63+2\kappa a)}{30\kappa} \label{eq:c1}
\end{align}
and 
\begin{equation}
c_2=\frac{4\pi}{15}\int_a^\infty dr\, r^2H(r)^2=\frac{a^2(3+2\kappa a)\pi}{10\kappa}. \label{eq:c2}
\end{equation}

Interestingly, in $c_1$ and $c_2$, all terms containing exponentials and exponential integrals are canceled. This can be understood by applying the Lorentz reciprocal theorem again. This time, we take two isotropic systems with inverse screening lengths $\kappa+\delta$ and $\kappa$, respectively, with all other conditions the same in both systems.
Then,
\begin{align}
&{\bf U}\cdot[{\bf F}_0(\kappa+\delta,a)-{\bf F}_0(\kappa,a)]\nonumber\\
&=(2\kappa+\delta)\eta\delta\int_{r>a} dV\, {\bf v}_0({\bf r};\kappa+\delta,a)\cdot{\bf v}_0({\bf r};\kappa,a),
\end{align}
where we write the explicit parameters as an argument to distinguish what isotropic system is used. Now take ${\delta\rightarrow 0}$ and we find
\begin{gather}
{\bf U}\cdot\frac{\partial}{\partial\kappa}{\bf F}_0(\kappa,a)=2\kappa\eta\int_{r>a}dV\, {\bf v}_0({\bf r};\kappa,a)\cdot{\bf v}_0({\bf r};\kappa,a).
\end{gather}
However, ${\bf F}_0(\kappa,a)=\zeta_0(\kappa,a){\bf U}$ and using the parametrization of Eq.~\eqref{eq:dyad}, we find the exact relation
\begin{equation}
\frac{\partial}{\partial\kappa}\zeta_0(\kappa,a)=2\kappa\eta[c_1(\kappa,a)+3c_2(\kappa,a)].
\end{equation}
The left-hand side, determined by using Eq.~\eqref{eq:isofric}, is consistent with the values found for the coefficients Eqs.~\eqref{eq:c1} and \eqref{eq:c2}.
\bibliography{literature1} 
\end{document}